\documentclass[12pt]{article}
\usepackage{amsmath}
\usepackage{graphicx}
\usepackage{eufrak}
\newlength{\dummysp}
\newcommand{\diag}{\mathop{{\hbox{diag} \, }}\nolimits}
\newcommand{\tr}{\mathop{{\hbox{Tr} \, }}\nolimits}

\newcommand{\stxt}[1]{\mathop{\hbox{{\scriptsize #1}}}\nolimits}
\newcommand{\bbar}[1]{{\overline{#1}}}

\newcommand{\beq}{\begin{eqnarray}}
\newcommand{\eeq}{\end{eqnarray}}
\newcommand{\nnn}{ \nonumber \\ }

\newcommand{\Zbf}{{{\bf Z}}}

\newcommand{\e}{{\epsilon}}
\newcommand{\s}{{\sigma}}
\newcommand{\vev}[1]{{\langle #1 \rangle}}
\newcommand{\bigvev}[1]{{\left\langle #1 \right\rangle}}
\newcommand{\ord}[1]{{{\cal O}(#1)}}
\newcommand{\gappeq}{\mathrel{\rlap {\raise.5ex\hbox{$>$}}
{\lower.5ex\hbox{$\sim$}}}}
\newcommand{\lappeq}{\mathrel{\rlap{\raise.5ex\hbox{$<$}}
{\lower.5ex\hbox{$\sim$}}}}
\newcommand{\myref}[1]{(\ref{#1})}

\newcommand{\ite}[1]{\vspace{5pt} {\it #1 \hspace{2pt}}}

\newcommand{\ben}{\begin{enumerate}}
\newcommand{\een}{\end{enumerate}}

\newcommand{\sqtw}{\sqrt{2}}

\newcommand{\sbar}{{\bar \s}}

\newcommand{\psib}{{\bar \psi}}

\newcommand{\bit}{\begin{itemize}}
\newcommand{\eit}{\end{itemize}}
\newcommand{\obf}{{\bf 1}}
\newcommand{\nbf}{{\bf n}}
\newcommand{\mbf}{{\bf m}}
\newcommand{\xb}{{\bbar{x}}}
\newcommand{\yb}{{\bbar{y}}}
\newcommand{\ibf}{\boldsymbol{\hat \imath}}
\newcommand{\jbf}{\boldsymbol{\hat \jmath}}
\newcommand{\sbf}{\boldsymbol{\s}}

\def\[{\left [}
\def\]{\right ]}
\def\({\left (}
\def\){\right )}

\begin{document}

\begin{titlepage}

\renewcommand{\thefootnote}{\fnsymbol{footnote}}

\hfill July 2, 2003

\hfill hep-lat/0304006

\vspace{0.25in}

\begin{center}
{\bf \Large Non-positive fermion determinants \\
\vskip 5pt
in lattice supersymmetry}
\end{center}

\vspace{0.15in}

\begin{center}
{\bf \large Joel Giedt\footnote{{\tt giedt@physics.utoronto.ca}}}
\end{center}

\vspace{0.15in}

\begin{center}
{\it University of Toronto \\
60 St. George St., Toronto ON M5S 1A7 Canada}
\end{center}

\vspace{0.15in}

\begin{abstract}
We find that fermion determinants are not generally
positive in a recent class of constructions
with explicit lattice supersymmetry.
These involve an orbifold of supersymmetric matrix models,
and have as their target (continuum) theory
(2,2) 2-dimensional super-Yang-Mills.
The fermion determinant is shown to be identically zero for all
boson configurations due to the existence of a zeromode
fermion inherited from the ``mother theory.''
Once this eigenvalue is factored out, the
fermion determinant generically has arbitrary complex
phase.  We discuss the implications of this
result for simulation of the models.

\end{abstract}

\end{titlepage}

\renewcommand{\thefootnote}{\arabic{footnote}}
\setcounter{footnote}{0}

\ite{Introductory remarks.}
Models with {\it exact} lattice supersymmetry have
been discussed in the literature by a few
groups.  For example, latticizations of super-Yang-Mills
\cite{Cohen:2003xe,Kaplan:2002wv}, supersymmetric
quantum mechanics \cite{Catterall:2000rv,Catterall:2001wx},
the 2d Wess-Zumino model \cite{Catterall:2001fr,Catterall:2001wx},
and direct constructions in the spirit
of the Ginsparg-Wilson relation---as suggested by L\"uscher \cite{Luscher:1998pq}
and worked out in {\it noninteracting}
examples \cite{Aoyama:1998in,Bietenholz:1998qq}---have
all been considered.\footnote{An approach very similar
to \cite{Bietenholz:1998qq} has been applied
in \cite{So:1998ya}, yielding slightly different
expressions.}
In this letter we will be interested in the
super-Yang-Mills constructions that lead to
a Euclidean lattice theory \cite{Cohen:2003xe}.
The method of building such models is based on {\it deconstruction}
of extra dimensions \cite{Arkani-Hamed:2001ca,Hill:2000mu}.  The
corresponding interpretation in terms of the world-volume
theory of D-branes has led to the latticizations of
2d, 3d and 4d supersymmetric gauge theories.
These lattice constructions are all arrived at by {\it orbifold projections}
of supersymmetric matrix models; i.e., in each case we quotient a
matrix model by some discrete symmetry group of the
theory.  Degrees of freedom that are not
invariant with respect to the combined action
of the orbifold generators are projected out.\footnote{For
a detailed discussion, we refer the reader to \cite{Cohen:2003xe}.}
Thus, throughout this letter we will have occasion to
speak of ``orbifolded'' matrix models and ``nonorbifolded''
matrix models.

A major motivation for efforts to latticize supersymmetric
models is that some
nonperturbative aspects of supersymmetric field
theories are not accessible by the usual techniques,
such as holomorphy.  One hope of a lattice supersymmetry
program of research is that it would lead to, e.g., simulations
that would provide further data on supersymmetric
field theories, especially those that include
super-Yang-Mills.\footnote{For a recent
review of existing work on this broad topic,
and a complete list of relevant references,
see \cite{Feo:2002yi}.}  With exact
lattice supersymmetry the target (continuum)
theory may be obtained in a more controlled fashion.
Indeed, in some cases it may be obtained without
the need for fine-tuning \cite{Cohen:2003xe,Kaplan:2002wv}.

In this enterprise, it is of great practical
importance that the fermion determinant, obtained
integrating over the fermion degrees of freedom
in the partition function, be positive.  For let
$\phi$ be the lattice bosons in the theory,
and $\psi,\psib$ the lattice fermions.  We obtain\footnote{
Here we assume ordinary finite dimensional Berezin
integration for the fermion coordinates; thus we
avoid subtleties of the {\it Weyl determinant}
present in continuum theories
\cite{Alvarez-Gaume:1983ig,Hsu:1997df}.  The class of
theories studed here result from a 4d $\to$ 0d dimensional
reduction, where the Pfaffian one obtains has an
equivalent determinant form \cite{Krauth:1998xh}.}
\beq
Z &=& \int [d \phi d \psib d \psi] \exp \[-S_B(\phi)-\psib M(\phi) \psi\] \nnn
&=& \int [d \phi] \det M(\phi) \exp \[-S_B(\phi)\] \nnn
&=& \int [d \phi] \exp \[ -S_{\stxt{eff}}(\phi) \] , \\
S_{\stxt{eff}}(\phi) &=& S_B(\phi) - \ln \det M(\phi).
\label{sefd}
\eeq
A positive\footnote{Strictly speaking,
we require only positive semi-definite $\det M(\phi)$,
provided $\det M(\phi)=0$ occurs only for
some subset of the boson configurations; this
subset has measure zero.  However, if $\det M(\phi)\equiv 0$,
i.e., for all boson configurations, then $Z$ is not well-defined.
This latter situation is what occurs in the theory that we study here.}
$\det M(\phi)$ for all $\phi$ allows us to unambiguously
calculate in an equivalent bosonic theory with action
$S_{\stxt{eff}}(\phi)$.
Expectation values of operators $\ord{\phi}$ may be obtained using
this action:
\beq
\vev{\ord{\phi}} = \frac{\int [d \phi] \ord{\phi} \exp \[ -S_{eff}(\phi) \]}
{\int [d \phi] \exp \[ -S_{eff}(\phi) \]}
\label{opd}
\eeq
Fermionic correlators will simply involve $M^{-1}(\phi)$
in the operator of interest.
In the case of positive
$\det M(\phi)$,
the techniques of estimation by Monte Carlo simulation
are robust; otherwise the situation is murkier.

\vspace{5pt}

We now summarize the content of our work:
\bit
\item
In this letter we show that the lattice
theory with (2,2) 2d super-Yang-Mills
as its target \cite{Cohen:2003xe}, obtained from
orbifolded supersymmetric matrix models, possesses a problematic
fermion determinant.  Due to a zeromode fermion,
$\det M(\phi)\equiv 0$, i.e., for all boson configurations.
\item
We then suggest how the zero eigenvalue can be factored
out in a controlled way in order to exhibit the
determinant for the other fermions.
For the orbifolded matrix models, we carry out
this factorization (numerically) in the special case of a $U(2)$
gauge theory, and show that it is robust.
As an example, we present our results for a $2 \times 2$ lattice.
\item
We further validate our method by applying it
to nonorbifolded $U(k)$ supersymmetric matrix models, which
also contain ever-present zeromode fermions,
corresponding to the $U(1)$ ``gaugino'' in the decomposition
$U(k) \supset U(1) \times SU(k)$.
We correctly reproduce the fermion determinant
for the nonorbifolded
$SU(k)$ matrix models by our factorization method.
As an example, we present our results for
$U(5) \supset U(1) \times SU(5)$.
\item
Once the zeromode fermion has been factored out of the
orbifolded matrix models studied here,
{\it we find that the remaining product of eigenvalues
is generically nonzero with arbitrary complex phase.}
\item
We explain how this is not in conflict with
results in nonorbifolded $SU(k)$ supersymmetric matrix models.
(For example, in an appendix of \cite{Ambjorn:2000bf}
it was shown that the product of eigenvalues
for the fermion matrix is positive semi-definite.)
It is found that the orbifold procedure lies at the
heart of this matter.
\item
We conclude with a discussion of the implications of
out results for lattice simulations of the
latticized (2,2) 2d super-Yang-Mills theories.
\eit

\ite{Zeromode fermion.}
Here we will focus on the orbifolded supersymmetric
matrix models that have as their target
theory (2,2) 2d super-Yang-Mills with $U(k)$ gauge group.  The
``mother theory'' is a nonorbifolded $U(kN^2)$ supersymmetric
matrix model.  The ``daughter theory'' is obtained by
orbifolding the ``mother theory'' by a $Z_N \times Z_N$
symmetry group, leaving intact (among other things) a $U(k)^{N^2}$
symmetry group that will become the gauge symmetry
of the $N \times N$ lattice theory.  The lattice theory is obtained
by studying the ``daughter theory'' about a particular
boson configuration that is a stationary point
of the action; i.e., it is a point in the moduli
space of the ``daughter theory.''

The orbifolded matrix model discussed here has been described in
detail in \cite{Cohen:2003xe}; we refer the reader there for further
details.  For our purpose it suffices to note that
in the $U(k)$ case the theory contains\footnote{Our
notation is standard:  $\mbf$ labels sites on a 2d
square lattice, so that $\mbf \in \Zbf \times \Zbf$, with
$\Zbf$ the set of integers.}
bosons $x_\mbf,y_\mbf$ that are $k \times k$ complex
matrices.  The bosons $x_\mbf,y_\mbf$ may written in terms of a
Hermitian basis
\beq
T^\mu \in \left\{ \sqrt{\frac{2}{k}} \obf_k, T^a \right\},
\qquad (T^a)^\dagger = T^a, \qquad
x_\mbf = x_\mbf^\mu T^\mu, \qquad
y_\mbf = y_\mbf^\mu T^\mu.
\eeq
It is always possible to choose the $T^a$ such that
\beq
\tr(T^\mu T^\nu) = 2 \delta^{\mu \nu} .
\eeq
Furthermore we define
\beq
\tr (T^\mu T^\nu T^\rho) = 2 \sqrt{\frac{2}{k}}~ t^{\mu \nu \rho}
\eeq
and note that (underlining implies all permutations are to
be taken):
\beq
t^{\underline{\mu \rho 0}} = \delta^{\mu \rho}.
\eeq
One finds that the fermionic part of action is given by
\beq
S_F = \frac{2}{g^2 \sqrt{k}}
(\alpha_\mbf^\mu, \beta_\mbf^\mu) \cdot M_{\mbf,\nbf}^{\mu \rho}
\cdot \binom{\lambda_\nbf^\rho}{\xi_\nbf^\rho } ~ .
\label{sfdf}
\eeq
Here $\alpha_\mbf^\mu, \beta_\mbf^\mu,\lambda_\nbf^\rho,\xi_\nbf^\rho$
are the lattice fermions, with upper index corresponding to the basis
$T^\mu$ introduced above.  The fermion matrix $M_{\mbf,\nbf}^{\mu \rho}$
is given by (sum over $\nu$ implied in the entries,
$\ibf,\jbf$ unit vectors):
\beq
M_{\mbf,\nbf}^{\mu \rho} = \(
\begin{array}{c|c}
t_{\mbf, \nbf}^{\mu \nu \rho} \xb_\mbf^\nu
- t_{\mbf, \nbf-\ibf}^{\mu \rho \nu} \xb_\mbf^\nu &
- t_{\mbf, \nbf}^{\mu \nu \rho} y_{\mbf+\ibf}^\nu
+ t_{\mbf, \nbf+\jbf}^{\mu \rho \nu} y_\nbf^\nu \\ \hline
t_{\mbf, \nbf}^{\mu \nu \rho} \yb_\mbf^\nu
- t_{\mbf, \nbf-\jbf}^{\mu \rho \nu} \yb_\mbf^\nu &
t_{\mbf, \nbf}^{\mu \nu \rho} x_{\mbf+\jbf}^\nu
- t_{\mbf, \nbf+\ibf}^{\mu \rho \nu} x_\nbf^\nu
\end{array}
\) ~ .
\label{kjtw}
\eeq
Here we have introduced the compact notation
\beq
t_{\mbf, \nbf}^{\mu \nu \rho} = \delta_{\mbf,\nbf} t^{\mu \nu \rho}.
\eeq
The fermion zero mode is easily established.  We consider
fermions of the form
\beq
\binom{\lambda_\nbf^\rho}{\xi_\nbf^\rho }
= \binom{\delta^{\rho 0} \lambda}{0},
\qquad \forall \; \nbf.
\label{zmf}
\eeq
Then in this case
\beq
&& \sum_\nbf \xb_\mbf^\nu (t_{\mbf, \nbf}^{\mu \nu \rho}
- t_{\mbf, \nbf-\ibf}^{\mu \rho \nu})
\lambda_\nbf^\rho = \sum_\nbf \xb_\mbf^\nu (t_{\mbf, \nbf}^{\mu \nu 0}
- t_{\mbf, \nbf-\ibf}^{\mu 0 \nu}) \lambda \nnn
&& \qquad = \xb_\mbf^\mu \lambda
\sum_\nbf (\delta_{\mbf,\nbf} - \delta_{\mbf,\nbf-\ibf}) =0.
\eeq
By similar arguments
\beq
\sum_\nbf \yb_\mbf^\nu (t_{\mbf, \nbf}^{\mu \nu \rho}
- t_{\mbf, \nbf-\jbf}^{\mu \rho \nu}) \lambda_\nbf^\rho
= 0.
\eeq
Thus \myref{zmf} is an eigenvector of $M$ with eigenvalue zero.
It follows that $\det M(x,y) \equiv 0$, for all configurations
of bosons.  This clearly poses a difficulty for
defining $S_{\stxt{eff}}(x,y)$.  We will shortly address a method
to factor out this zero eigenvalue, and study it in
some detail for the $U(2)$ case.
However, we first make a few remarks on the existence
of the indentically zero eigenvalue.

The zeromode fermion is nothing but the
``zero-momentum'' mode of the Fourier transform
of a given $\lambda_\mbf^0$:
\beq
\lambda \equiv \tilde \lambda_{\bf 0}^0 =
\frac{1}{N} \sum_{\mbf} \lambda_\mbf^0 .
\eeq
In the formalism introduced in \cite{Cohen:2003xe},
$\lambda_\mbf^0$ appears in the superfield
\beq
\Lambda^0_\mbf = \lambda_\mbf^0 - \[ \xb_{\mbf-\ibf}^\mu
x_{\mbf-\ibf}^\mu - \xb_{\mbf}^\mu x_{\mbf}^\mu
+ \yb_{\mbf-\jbf}^\mu
y_{\mbf-\jbf}^\mu - \yb_{\mbf}^\mu y_{\mbf}^\mu
+ id_\mbf^0 \] \theta .
\eeq
Here $d_\mbf^0$ is an auxiliary boson and $\theta$ is an odd
(Grassman) superspace coordinate.
The zero-momentum part of this supermultiplet is just
\beq
\Lambda \equiv \sum_\mbf \Lambda^0_\mbf
= \lambda + i d \theta, \qquad
d \equiv \frac{1}{N} \sum_\mbf d_\mbf^0 .
\eeq
That is, the zeromode fermion is in a multiplet that contains
just itself and an auxiliary boson.  It
is a lattice version of a
{\it Fermi multiplet} \cite{Witten:1993yc}.  Thus there is no {\it physical}
zeromode boson that corresponds to the zeromode fermion.

The existence of the zeromode fermion can be understood
in terms of the ``mother theory;'' i.e., the nonorbifolded
matrix model.  There, the bosons
$x,y$ and their conjugates are understood in terms of a ``vector boson''
$v$:
\beq
v = v_m \sbar_m = v_0 + i {\bf v \cdot} \sbf = \sqtw
\begin{pmatrix} \xb & -y \cr \yb & x \cr \end{pmatrix}
\eeq
where $v_m$ are Hermitian matrices
that are Lie algebra valued in $U(kN^2)$:
\beq
(v_m)_{i \mu \nu, j \mu' \nu'}
= v_m^\alpha (T^\alpha)_{i \mu \nu, j \mu' \nu'}
\qquad i,j = 1,\ldots,k, \quad
\mu,\nu,\mu',\nu' = 1,\ldots N.
\eeq
Note that the indices of generators have been written in such
a way that the $U(k)^{N^2}$ subgroup has been manifestly factored
out.  One should think of one $U(k)$ factor ``living''
at each site.  Similarly, the fermions of the mother theory are given by
\beq
\psi_{i \mu \nu, j \mu' \nu'} =
\psi^\alpha (T^\alpha)_{i \mu \nu, j \mu' \nu'}, \qquad
\psi^\alpha = \binom{\lambda^\alpha}{\xi^\alpha}
\eeq
with a corresponding expression for $\psib$.
The fermion action in the mother theory takes the form
\beq
S_F = \frac{1}{g^2} \tr \( \psib \sbar_m [v_m , \psi] \) .
\label{mmvd}
\eeq

Now note that the diagonal $U(1)_{\stxt{diag}}
\subset U(kN^2)$ fermions
do not appear $S_F$.  That is,
\beq
\psi^\alpha (T^\alpha)_{i \mu \nu, j \mu' \nu'}
\ni \psi^0 \delta_{i \mu \nu, j \mu' \nu'}
\label{sfhe}
\eeq
and $\psi^0$ disappears because of the commutator.  Since
$\psi^0$ is a two component fermion, it gives
two zeromode fermions of the mother theory, independent
of the boson configuration $v_m$.
The boson action takes the form
\beq
S_B = - \frac{1}{4g^2} \tr \( [v_m, v_n] [v_m, v_n] \) .
\eeq
Here again, the $U(1)_{\stxt{diag}}$ boson disappears from the
action because of the commutators.  It is a zeromode
for all boson configurations.

Most of the fields of the mother theory
are projected out in the orbifold
construction.  The projections depend on charges with
respect to a $U(1)_{r_1} \times U(1)_{r_2}$ global
symmetry group.  As it turns out, the $\lambda^\alpha$
fermions are $U(1)_{r_1} \times U(1)_{r_2}$ neutral.
It follows that only the diagonal parts
with respect to the ``site'' indices $(\mu \nu,\mu' \nu')$
survive in the orbifolded matrix model (``daughter theory''):
\beq
\lambda^\alpha (T^\alpha)_{i \mu \nu, j \mu' \nu'}
\to \lambda^\alpha (T^\alpha)_{i \mu \nu, j \mu \nu} ~ .
\eeq
But this includes the zeromode fermion of the mother theory.
On the other hand, none of the components of $v$ are
$U(1)_{r_1} \times U(1)_{r_2}$ neutral.  It follows
that, in the orbifolded matrix model,
they are all off-diagonal\footnote{In the lattice
theory, these are interpreted as link fields beginning
at the site labeled by $(\mu, \nu)$ and
ending at the site labeled by $(\mu', \nu')$.} with respect to
the indices $(\mu \nu,\mu' \nu')$.  Hence the zeromode bosons
of the mother theory are projected out in the orbifolded theory.
Similarly, none of the $\xi^\alpha$ fermions of the mother theory
are $U(1)_{r_1} \times U(1)_{r_2}$ neutral, so that
the second zeromode fermion $\xi^0$ of the mother
theory disappears in the projection.

\ite{Deformed $U(k)$.}
The zeromode eigenvalue of the orbifolded matrix
model can be factored out as follows.  We
deform the fermion matrix \myref{kjtw} according to
\beq
M \to  M + \e {\bf 1}_{N_f}
\eeq
where $N_f$ is the dimensionality of the fermion matrix
and $\e \ll 1$ is a deformation parameter that we will
eventually take to zero.  We factor out the zero mode
through the definition
\beq
&& \hat M(0) = \lim_{\e \to 0} \hat M(\e), \qquad
\hat M(\e) = \e^{-1/N_f} (M + \e {\bf 1}_{N_f} ), \nnn
&& \qquad \Rightarrow \quad
\det \hat M(0) = \lim_{\e \to 0} \e^{-1} \det (M + \e {\bf 1}_{N_f} ) .
\eeq

If this deformation is added to the action, it explicitly
breaks the exact lattice supersymmetry and gauge invariance.
This infrared regulator could be removed in the continuum
limit, say, by taking $\epsilon a \ll N^{-1}$.  Noting that
$L = Na$ is the physical size of the lattice, the
equivalent requirement is that $\epsilon \ll L^{-1}$
be maintained as $a \to 0$, for fixed $L$.  Thus in the
thermodynamic limit ($L \to \infty$), the deformation is
removed.  The parameter $\epsilon$ is a soft
infrared regulating mass, and is quite analogous to the
soft mass $\mu$ introduced by Cohen et al.~\cite{Cohen:2003xe}
in their Eq.~(1.2) to control the bosonic zeromode of
the theory.  In the same way that a finite $\mu$
does not modify the final results of the renormalization
analysis of Section 5 of \cite{Cohen:2003xe}, our $\epsilon$
does not modify the result of the quantum continuum limit.
The essence of the argument is that we have introduced a
vertex that will be proportional to the dimensionless
quantity $g_2^2 \epsilon a^3 \ll g_2^2 a^3 / L$, where
$g_2$ is the 2d coupling constant.  Such
contributions to the operator coefficients $C_{{\cal O}}$
in Eq.~(5.2) of \cite{Cohen:2003xe} vanish in the thermodynamic
limit and because the target theory is super-renormalizable,
we are assured that the perturbative power counting
arguments are reliable and the correct continuum limit is
obtained.

Of course it remains to study the convergence of
$\det \hat M(\e) \to \det \hat M(0)$.  Indeed, for
$U(2)$ lattice theory (i.e., the orbifolded matrix model
with $U(2)^{N^2}$ symmetry) we have performed
this analysis numerically for a large number of boson
configurations drawn randomly from a Gaussian distribution,
as will be detailed below.  We find that the convergence
is rapid and that a reliable estimate for $\det \hat M(0)$
can be obtained in this way.  Furthermore, we find that
for $\e \ll 1$, the phase of $\det \hat M(\e)$ quickly
converges to a constant value, and that it is
uniformly distributed throughout the interval $(-\pi,\pi]$
for the random Gaussian boson variables.

As a check of our method, we have studied also the
analogous deformation of the nonorbifolded $U(k)$ supersymmetric
matrix models, where the two zeromode fermions in $\psi^0$ are
present.  Indeed, as will be discussed below,
we find that
\beq
\lim_{\e \to 0} \e^{-2} \det (M_{U(k)} + \e {\bf 1}_{N_f} )
= \det M_{SU(k)} .
\label{dfmm}
\eeq
That is, we obtain the determinant of the nonorbifolded
$SU(k)$ supersymmetric matrix model, which has the zeromode
fermions factored out.  With appropriate conventions,
$\det M_{SU(k)}$ is positive semi-definite.\footnote{For
some conventions on the $SU(k)$ generators and the
overall coefficient in $S_F$ of Eq. \myref{mmvd} it is possible
that a constant phase (i.e., independent of boson configurations)
may be present.  However, this factors out of \myref{opd} and is
of no concern.}

\ite{Deformed $U(2)$.}
Here we specialize to the orbifolded matrix model
with $U(2)^{N^2}$ symmetry, which becomes the $U(2)$
gauge invariance of an $N \times N$ lattice theory.
In this case we take
\beq
T^\mu \in \{ \obf_2, \s^a \}, \qquad
x_\mbf = x_\mbf^0 + x_\mbf^a \s^a, \qquad
y_\mbf = y_\mbf^0 + y_\mbf^a \s^a.
\label{qqre}
\eeq
Then the fermion matrix is given in \myref{kjtw}, where
in the present case
\beq
t^{000}=1, \qquad t^{\underline{a00}} = 0, \qquad
t^{\underline{ab0}} = \delta^{ab}, \qquad t^{abc} = i\e^{abc} .
\eeq
The lattice theory is obtained by expansion about
a point in moduli space:
\beq
x_\mbf^0 = \frac{1}{a\sqtw} + \cdots, \qquad
y_\mbf^0 = \frac{1}{a\sqtw} + \cdots, \qquad
\eeq
where $\cdots$ represent the quantum fluctuations.
For this reason, in our study of $\det \hat M(\e)$
we scan over a Gaussian distribution
where $x_\mbf^0,y_\mbf^0$ have a a nonzero mean $1/a\sqtw \equiv 1$.
The remainder of the bosons are drawn with mean
zero.  All bosons are taken from distributions
with unit variance.

\begin{figure}
\begin{center}
\includegraphics[height=5.0in,width=3.0in,angle=90]{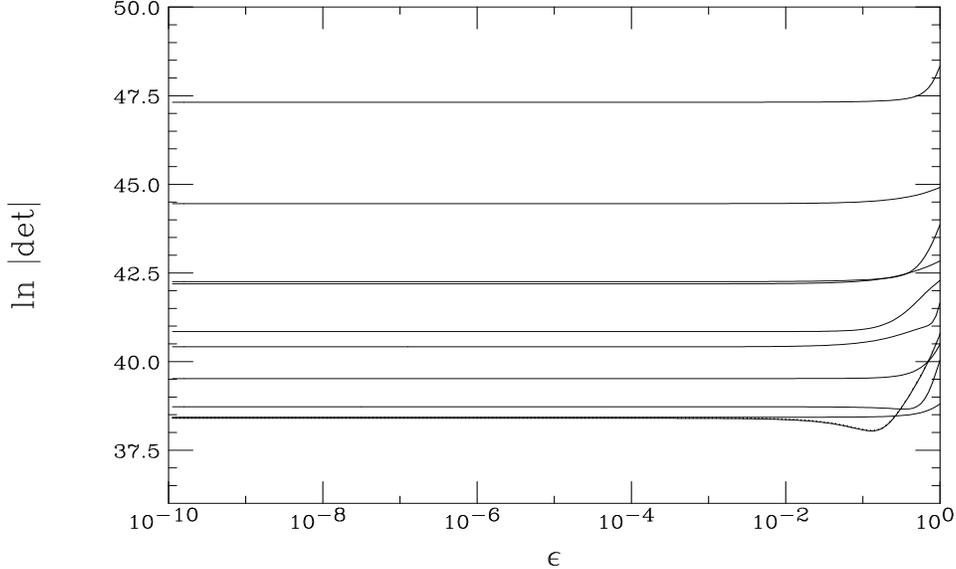}
\end{center}
\caption{$\ln |\det \hat M(\e)|$ versus $\e$ for a
sequence of random draws.  These results are for
the $U(2)$ lattice theory, with $2 \times 2$ lattice.}
\label{f1}
\end{figure}

\begin{figure}
\begin{center}
\includegraphics[height=5.0in,width=3.0in,angle=90]{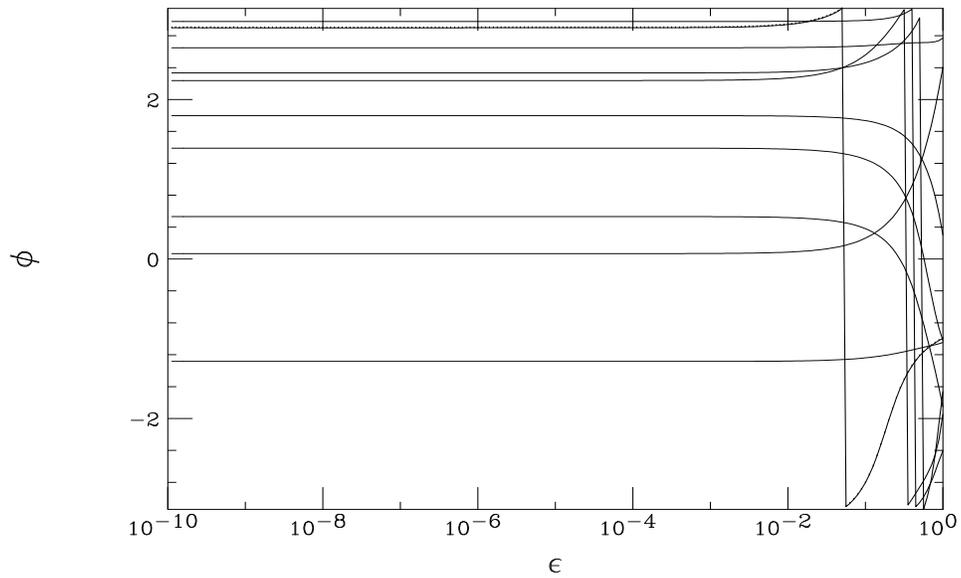}
\end{center}
\caption{$\phi = \arg \det \hat M(\e)$ versus $\e$ for a
sequence of random draws.
Note that crossing over the boundary of the domain $(-\pi,\pi]$
to an equivalent point within that domain is indicated by the nearly
vertical lines.  These results are for
the $U(2)$ lattice theory, with $2 \times 2$ lattice.}
\label{f2}
\end{figure}

In Figs. \ref{f1} and \ref{f2} we display $\ln |\det \hat M(\e)|$ and $\arg
\det \hat M(\e)$ versus $\e$ for the case of $N = 2$,
the smallest lattice possible, of size $2 \times 2$.  Each line corresponds
to a different random draw.  It can be seen that the convergence
is rapid and reliable.
As a check, we have computed
the eigenvalues of the undeformed matrix $M$,
using the math package Maple, for
the same set of random boson configurations.
We find that in each case the product of nonzero eigenvalues
agrees with $\det \hat M(0)$ in magnitude and phase
to within at least 5 significant digits.
This conclusively demonstrates that the complex determinant
is not an artifact of the deformation, but is a property of
the nonzero eigenvalues of the undeformed matrix $M$.

\begin{figure}
\begin{center}
\includegraphics[height=5.0in,width=3.0in,angle=90]{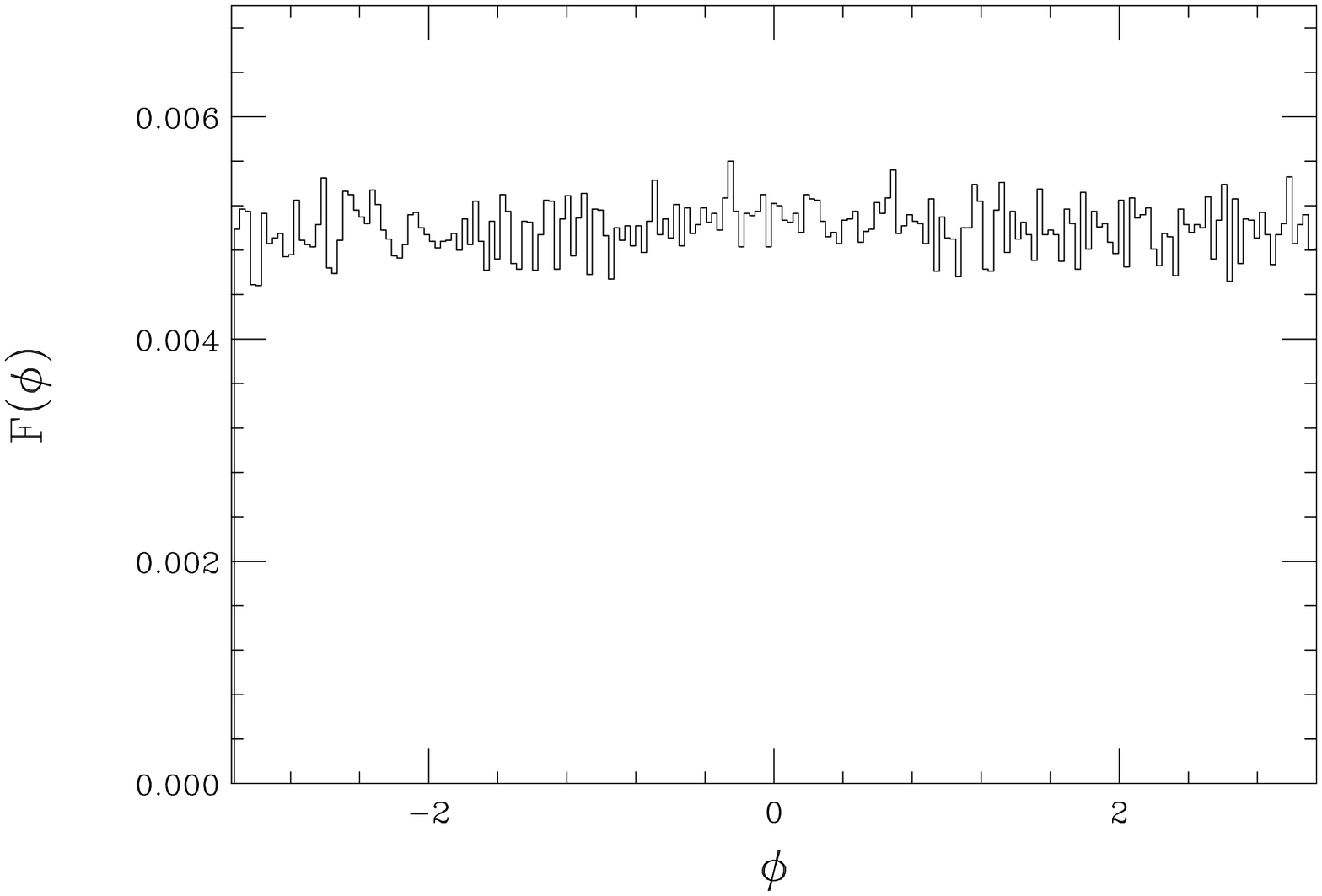}
\end{center}
\caption{Frequency distribution $F(\phi)$
for $\phi = \arg \det \hat M(0)$,
for $10^5$ random (Gaussian) draws, binned into intervals of $\pi/100$.
The distribution of $\phi$ is seen to be nearly uniform.  These results are for
the $U(2)$ lattice theory, with $2 \times 2$ lattice.}
\label{f3}
\end{figure}

For a set of $10^5$ draws on the bosons of this
$2 \times 2$ lattice, we have extrapolated to $\e \to 0$ and
binned $\arg \det \hat M(\e)$ over its range, with
bins of size $\pi/100$.  In Fig. \ref{f3}
we show the frequency for each bin, as a fraction of
the total number of draws.  In the extrapolation, we decreased
$\e$ by powers of 10 until the change in both
$\ln |\det \hat M(\e)|$ and $\arg \det \hat M(\e)$
were both less than $10^{-4}$ per decade.  This was
done for each draw to get a reliable estimate for
$\det \hat M(0)$.  In each of the $10^5$ draws the
$10^{-4}$ per decade criterion was reached well before
$\e = 10^{-10}$.

To summarize, once the zeromode eigenvalue is factored out,
the product of the nonzero eigenvalues has arbitrary phase.
Consequently an ambiguity exists in defining $S_{\stxt{eff}}(x,y)$
for the orbifolded matrix model theory.
We are presently exploring whether or
not this can be overcome for the purposes of simulation.
At present, however, all we can say is that
this difficulty is rather troubling.

\ite{Comparison to nonorbifolded $U(k)$ and $SU(k)$ matrix models.}
As mentioned above, the nonorbifolded
supersymmetric $U(k)$ matrix models contain two ever-present
zeromode fermions in the $\psi^0$ that appears in \myref{sfhe}.
Then $\det M_{U(k)} \equiv 0$.  However,
it is easy enough to just work with the nonorbifolded $SU(k)$ matrix
model, so that $\psi^0,\psib^0$ are never in the theory to
begin with.  Then with appropriate conventions $\det M_{SU(k)}
\ge 0$.  A proof of this result has been given in
an appendix of \cite{Ambjorn:2000bf}.

\begin{figure}
\begin{center}
\includegraphics[height=5.0in,width=3.0in,angle=90]{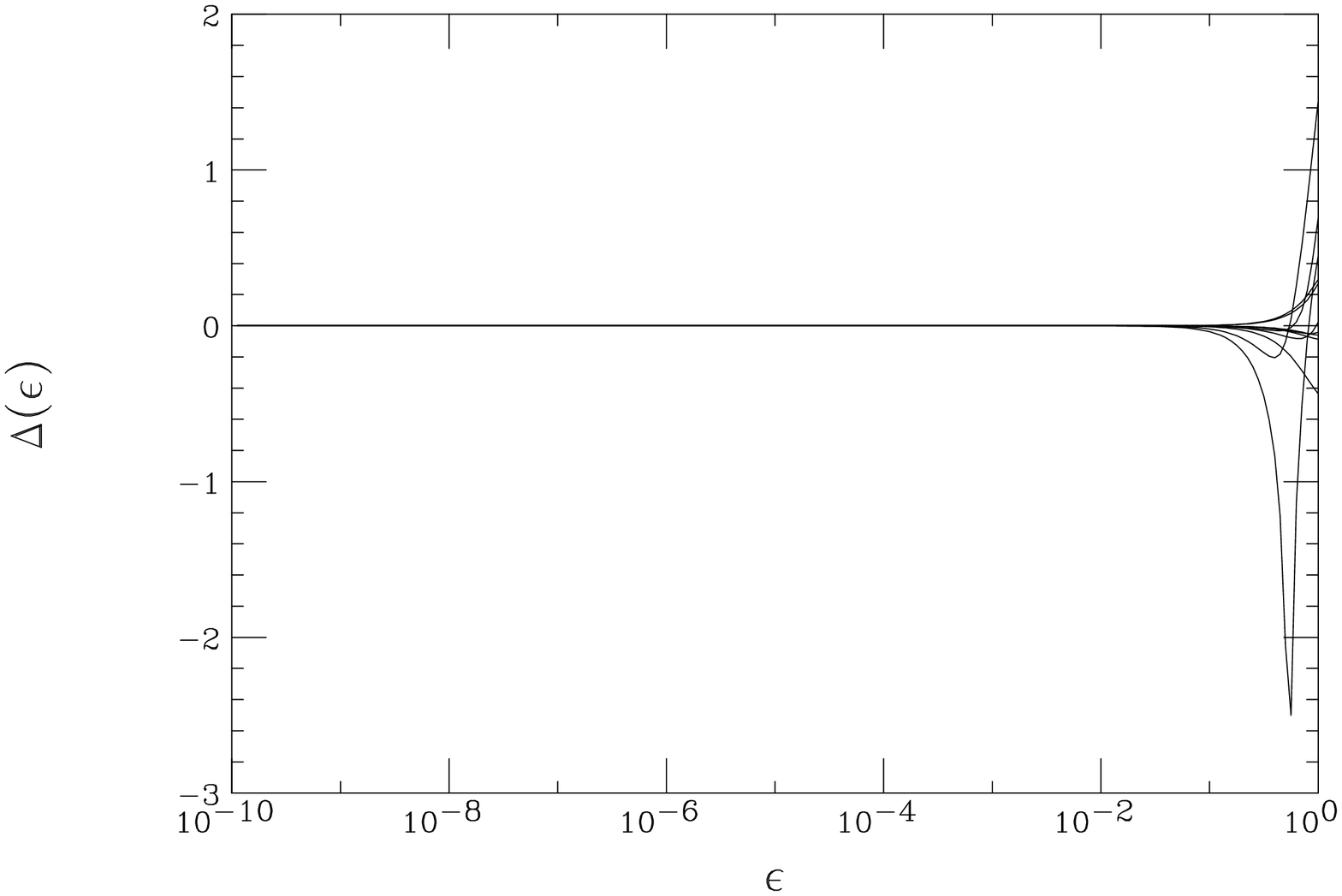}
\end{center}
\caption{$\Delta(\e)$ versus $\e$ for a sequence of
random draws in the nonorbifolded $U(5)$ matrix model.}
\label{fm1}
\end{figure}

\begin{figure}
\begin{center}
\includegraphics[height=5.0in,width=3.0in,angle=90]{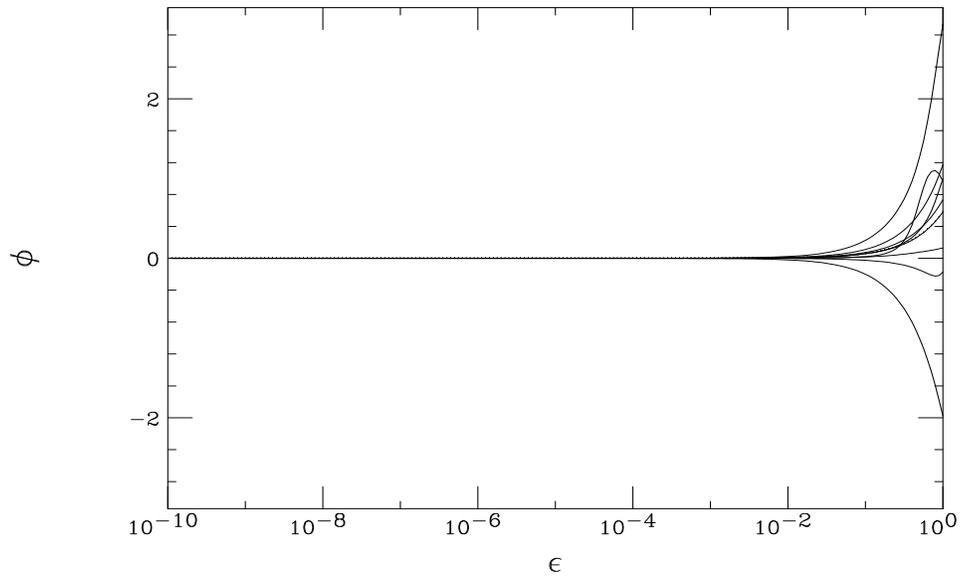}
\end{center}
\caption{$\phi$ versus $\e$ for sequence of random draws
in the nonorbifolded $U(5)$ matrix model.}
\label{fm2}
\end{figure}

As a test of our method, we have verified \myref{dfmm}
numerically, for a sequence of random Gaussian
draws on the bosons $v_m^\alpha$ that appear in
\myref{mmvd}.  In Fig. \ref{fm1} we show the quantity
\beq
\Delta(\e) =  \ln \[ \e^{-2} |\det (M_{U(k)} + \e {\bf 1}_{N_f} )| \]
- \ln |\det M_{SU(k)}|
\eeq
as a function of $\e$ for the case of $k=5$.  Indeed it can be
seen that the convergence as $\e \to 0$ is quite rapid.  We find that
$|\Delta(10^{-3})| \lappeq 10^{-4}$ and
$|\Delta(10^{-4})| \lappeq 10^{-9}$ as a rule.
In Fig. \ref{fm2} we show the quantity
\beq
\phi(\e) =  \arg \[ \e^{-2} \det (M_{U(k)} + \e {\bf 1}_{N_f} ) \]
\eeq
as a function of $\e$, again for $k=5$.  The value for
$SU(5)$ in our conventions is $\phi=0$.
It can be seen that a rapid convergence
to this value is obtained.

One might wonder how the generic phases of Fig. \ref{f3} are
obtained in the orbifolded matrix models, given the positivity
of the fermion determinant in the nonorbifolded $SU(k)$ matrix models.
Firstly, one should note that the proof given in
\cite{Ambjorn:2000bf} shows only that for each
eigenvalue $\mathfrak{e}$ of $M_{SU(k)}$ there exists
also an eigenvalue ${\mathfrak{e}}^*$, and that these
always come in pairs.  The proof relies essentially on
the relation
\beq
\s^2 v^\alpha \s^2 = \s^2 (v_0^\alpha + i {\bf v}^\alpha
{\bf \cdot} \sbf) \s^2 = (v^\alpha)^* .
\label{jrr}
\eeq
On the other hand, the $Z_N \times Z_N$ orbifold action
on the mother theory bosons is generated by
\beq
v^\alpha (T^\alpha)_{i \rho \nu, j \rho' \nu'} &\to&
e^{-i\pi \s^3/N} v^\alpha e^{-i\pi \s^3/N} ~
\Omega_{\rho \mu} (T^\alpha)_{i \mu \nu,j \mu' \nu'} \Omega^{-1}_{\mu' \rho'}
\nnn
v^\alpha (T^\alpha)_{i \rho \mu, j \rho' \mu'} &\to&
e^{i\pi \s^3/N} v^\alpha e^{-i\pi \s^3/N} ~
\Omega_{\rho \nu} (T^\alpha)_{i \mu \nu, j \mu' \nu'} \Omega^{-1}_{\nu' \rho'}
\eeq
where $\Omega=\diag(\omega,\omega^2,\ldots,\omega^N), \;
\omega=\exp(2\pi i/N)$.
Thus because of the $\exp(\pm i\pi \s^3/N)$ factors,
the orbifold projection does not commute with the
operations of the proof given in \cite{Ambjorn:2000bf};
i.e., Eq. \myref{jrr}.
The fermion matrix of the projected theory---i.e., the
orbifolded matrix model---lacks many
of the eigenvalues of the mother theory; it should come
as no surprise that not all eigenvalues are removed in
pairs $(\mathfrak{e},{\mathfrak{e}}^*)$.  After all,
we already know that only {\it one} of the zero eigenvalues
is removed from the $U(kN^2)$ mother theory.
For this reason it is not at all contradictory that the
projected theory has a product of nonzero eigenvalues
that is not positive, nor real.

\ite{Discussion.}
The existence of a zeromode fermion in the
orbifolded matrix models has been reliably handled
by our deformation method, and the approach is easily implemented
numerically.  Analytic methods in nonorbifolded
supersymmetric matrix
models have also involved deformations of the theory
to render quantities of interest well-defined
\cite{Moore:1998et}; however, these involve the introduction
of auxiliary fields, which are expensive to implement in
a simulation.  For this reason, we prefer the method described
here.  In spite of the complex fermion determinant in the
orbifolded matrix models,
our approach allows for a detailed study of the partition
function by Monte Carlo methods, analogous to what has
been performed in \cite{Krauth:1998xh} for
nonorbifolded supersymmetric
matrix models.  It would be interesting
to see how expectations based on continuum results might
be realized in the present context.  For example,
in \cite{Hsu:1997df} it was shown that 4d ${\cal N}=1$
pure super-Yang-Mills has a positive fermion determinant.
Since the (2,2) 2d target theory studied here is a dimensional
reduction of this model, it is rather surprising that
we find arbitrary complex phase.  However, it
remains to be seen whether or not the lattice action
described here possesses reflection positivity,
or whether this is only obtained in the continuum limit.
Study of this issue is in progress.

We suspect that the difficulties faced here may
be, broadly speaking, related to those faced in defining the
phase of the fermion measure in the attempts to realize
chiral gauge theories on the lattice with an exact
chiral gauge symmetry \cite{Luscher:1998du,Luscher:1999un,
Luscher:2000hn}.  In the present context, we have
an exact chiral fermionic symmetry:  lattice supersymmetry
acts on the $\lambda_\mbf,\xi_\mbf$
but not on the $\alpha_\mbf,\beta_\mbf$
that appear in \myref{sfdf}.  A resolution of one
problem may lead to answers for the other.

In our opinion, the deconstruction approach
to lattice supersymmetry remains an exciting topic,
whatever difficulties may face attempts to simulate the
theory.  For example, the relatively simple systems described here
provide another context in which the difficulties\footnote{See for example
\cite{Fujimura:cq,Gausterer:1992jz} and references therein.}
that plague the complex Langevin approach \cite{Parisi:cs,Karsch:1985cb}
and other complex action techniques might be studied.

We are presently exploring whether or
not the complex phase can be overcome for the purposes
of simulation.  A typical approach would be to compute
averages of an operator ${\cal O}$ from the {\it re-weighting}
identity:
\beq
\vev{ {\cal O} } = \frac{\bigvev{ {\cal O} e^{i \phi} }_{p.q.}}
{ \bigvev{ e^{i \phi} }_{p.q.}}
\eeq
Here, $\phi= \arg \det \hat M(0)$, as above,
and ``p.q.'' indicates phase-quenching:
expectation values are computed with the replacement $\det \hat M(0)
\to |\det \hat M(0)|$.  Thus the effective bosonic action
\beq
S'_{\stxt{eff}} = S_B - \ln |\det \hat M(0)|
\eeq
is used to generate the phase-quenched ensemble
by standard Monte Carlo techniques.
However, it is well-known that this suffers efficiency
problems:  the number of configurations required to get
an accurate estimate for, say,
$\bigvev{ \exp ( i \phi ) }_{p.q.}$,
grows like $\exp (\Delta F \cdot N_f^2)$.  Here,
$\Delta F$ is the difference in free energy densities between the
full ensemble and the phase-quenched ensemble.
Recall that $N_f$ is the dimensionality of the fermion
matrix.
On the other hand, if the phase-quenched distribution is highly
concentrated near one value of $\phi$, the phase essentially
factors out of the partition function and efficient
simulations can be done using the phase-quenched ensemble.
Thus it is of interest to study the distribution of $\phi$
in the phase-quenched ensemble rather than the Gaussian distribution
used here.  Research in this direction is currently in progress
and we hope to report on it soon.

\vspace{10pt}

\noindent
{\bf \large Acknowledgements}

\vspace{3pt}

\noindent
The author would like to thank Erich Poppitz for numerous helpful
discussions and suggestions.  Thanks are also owed to
Wolfgang Bietenholz for enlightening communications,
and to the referee who made a number of useful comments.
This work was supported
by the National Science and Engineering Research Council of Canada.


\vspace{10pt}

\begin{thebibliography}{99}

\bibitem{Cohen:2003xe}
A.~G.~Cohen, D.~B.~Kaplan, E.~Katz and M.~Unsal,
``Supersymmetry on a Euclidean spacetime lattice. I: A target theory with  four supercharges,''
arXiv:hep-lat/0302017.

\bibitem{Kaplan:2002wv}
D.~B.~Kaplan, E.~Katz and M.~Unsal,
``Supersymmetry on a spatial lattice,''
arXiv:hep-lat/0206019.

\bibitem{Catterall:2000rv}
S.~Catterall and E.~Gregory,
``A lattice path integral for supersymmetric quantum mechanics,''
Phys.\ Lett.\ B {\bf 487} (2000) 349
[arXiv:hep-lat/0006013].

\bibitem{Catterall:2001wx}
S.~Catterall and S.~Karamov,
``A two-dimensional lattice model with exact supersymmetry,''
Nucl.\ Phys.\ Proc.\ Suppl.\  {\bf 106} (2002) 935
[arXiv:hep-lat/0110071].

\bibitem{Catterall:2001fr}
S.~Catterall and S.~Karamov,
``Exact lattice supersymmetry: the two-dimensional N = 2 Wess-Zumino  model,''
Phys.\ Rev.\ D {\bf 65} (2002) 094501
[arXiv:hep-lat/0108024].

\bibitem{Luscher:1998pq}
M.~L\"uscher,
``Exact chiral symmetry on the lattice and the Ginsparg-Wilson relation,''
Phys.\ Lett.\ B {\bf 428} (1998) 342
[arXiv:hep-lat/9802011].

\bibitem{Aoyama:1998in}
T.~Aoyama and Y.~Kikukawa,
``Overlap formula for the chiral multiplet,''
Phys.\ Rev.\ D {\bf 59} (1999) 054507
[arXiv:hep-lat/9803016].

\bibitem{Bietenholz:1998qq}
W.~Bietenholz,
``Exact supersymmetry on the lattice,''
Mod.\ Phys.\ Lett.\ A {\bf 14} (1999) 51
[arXiv:hep-lat/9807010].

\bibitem{So:1998ya}
H.~So and N.~Ukita,
``Ginsparg-Wilson relation and lattice supersymmetry,''
Phys.\ Lett.\ B {\bf 457} (1999) 314
[arXiv:hep-lat/9812002].

\bibitem{Arkani-Hamed:2001ca}
N.~Arkani-Hamed, A.~G.~Cohen and H.~Georgi, ``(De)constructing
dimensions,'' Phys.\ Rev.\ Lett.\  {\bf 86}, 4757 (2001)
[arXiv:hep-th/0104005];

\bibitem{Hill:2000mu}
C.~T.~Hill, S.~Pokorski and J.~Wang, ``Gauge invariant effective
Lagrangian for Kaluza-Klein modes,'' Phys.\ Rev.\ D {\bf 64},
105005 (2001) [arXiv:hep-th/0104035].

\bibitem{Feo:2002yi}
A.~Feo, ``Supersymmetry on the lattice,'' arXiv:hep-lat/0210015.

\bibitem{Alvarez-Gaume:1983ig}
L.~Alvarez-Gaume and E.~Witten,
``Gravitational Anomalies,''
Nucl.\ Phys.\ B {\bf 234} (1984) 269.

\bibitem{Hsu:1997df}
S.~D.~Hsu,
``Gaugino determinant in supersymmetric Yang-Mills theory,''
Mod.\ Phys.\ Lett.\ A {\bf 13} (1998) 673
[arXiv:hep-th/9704149].

\bibitem{Krauth:1998xh}
W.~Krauth, H.~Nicolai and M.~Staudacher,
``Monte Carlo approach to M-theory,''
Phys.\ Lett.\ B {\bf 431} (1998) 31
[arXiv:hep-th/9803117].

\bibitem{Ambjorn:2000bf}
J.~Ambjorn, K.~N.~Anagnostopoulos, W.~Bietenholz, T.~Hotta and J.~Nishimura,
``Large N dynamics of dimensionally reduced 4D SU(N) super Yang-Mills  theory,''
JHEP {\bf 0007} (2000) 013
[arXiv:hep-th/0003208].

\bibitem{Witten:1993yc}
E.~Witten,
``Phases of N = 2 theories in two dimensions,''
Nucl.\ Phys.\ B {\bf 403} (1993) 159
[arXiv:hep-th/9301042].

\bibitem{Moore:1998et}
G.~W.~Moore, N.~Nekrasov and S.~Shatashvili,
``D-particle bound states and generalized instantons,''
Commun.\ Math.\ Phys.\  {\bf 209} (2000) 77
[arXiv:hep-th/9803265].

\bibitem{Luscher:1998du}
M.~L\"uscher,
``Abelian chiral gauge theories on the lattice with exact gauge  invariance,''
Nucl.\ Phys.\ B {\bf 549} (1999) 295
[arXiv:hep-lat/9811032].

\bibitem{Luscher:1999un}
M.~L\"uscher,
``Weyl fermions on the lattice and the non-abelian gauge anomaly,''
Nucl.\ Phys.\ B {\bf 568} (2000) 162
[arXiv:hep-lat/9904009].

\bibitem{Luscher:2000hn}
M.~L\"uscher,
``Chiral gauge theories revisited,''
arXiv:hep-th/0102028.

\bibitem{Fujimura:cq}
K.~Fujimura, K.~Okano, L.~Schulke, K.~Yamagishi and B.~Zheng,
``On The Segregation Phenomenon In Complex Langevin Simulation,''
Nucl.\ Phys.\ B {\bf 424} (1994) 675
[arXiv:hep-th/9311174].

\bibitem{Gausterer:1992jz}
H.~Gausterer and S.~Lee,
``The Mechanism of complex Langevin simulations,''
arXiv:hep-lat/9211050.

\bibitem{Parisi:cs}
G.~Parisi,
``On Complex Probabilities,''
Phys.\ Lett.\ B {\bf 131} (1983) 393.

\bibitem{Karsch:1985cb}
F.~Karsch and H.~W.~Wyld,
``Complex Langevin Simulation Of The SU(3) Spin Model With Nonzero Chemical Potential,''
Phys.\ Rev.\ Lett.\  {\bf 55} (1985) 2242.

\end{thebibliography}
\end{document}